\documentclass[aps,prc,twocolumn,superscriptaddress]{revtex4}
\usepackage{graphicx}
\usepackage[section]{placeins}
\usepackage{float}
\usepackage[utf8]{inputenc}
\usepackage{tabularx}
\usepackage{booktabs}
\usepackage{threeparttable}
\usepackage{multirow}
\usepackage{color}
\usepackage{transparent}
\usepackage{epstopdf}
\usepackage{upgreek}
\usepackage{siunitx}
\usepackage{braket}
\usepackage{amsmath}
\usepackage{verbatim}

\newcommand{\beq}{\begin{equation}}
\newcommand{\eeq}{\end{equation}}
\newcommand{\myuparrow}{\mathord{\uparrow}}

\graphicspath{{../Pics/}}

\begin{document}

%---- Titel, Autoren und Infos  ----%
\title{Search for weak $M1$ transitions in $^{48}$Ca with inelastic proton scattering}

\newcommand{\TWMU}{Tokyo Women's Medical University, 8-1 Kawada-cho, Shinjuku-ku, Tokyo 162-8666, Japan}
\newcommand{\RCNP}{Research Center for Nuclear Physics, Osaka University, Ibaraki, Osaka 567-0047, Japan}
\newcommand{\TUDarmstadt}{Institut f\"ur Kernphysik, Technische Universit\"{a}t Darmstadt, D-64289 Darmstadt, Germany}

\author{M.~Mathy}\affiliation{\TUDarmstadt}
\author{J.~Birkhan}\affiliation{\TUDarmstadt}
\author{H.~Matsubara}\affiliation{\RCNP}\affiliation{\TWMU}
\author{P.~von~Neumann-Cosel}\email{vnc@ikp.tu-darmstadt.de}\affiliation{\TUDarmstadt}
\author{N.~Pietralla}\affiliation{\TUDarmstadt}
\author{V.~Yu.~Ponomarev}\affiliation{\TUDarmstadt}
\author{A.~Richter}\affiliation{\TUDarmstadt}
\author{A.~Tamii}\affiliation{\RCNP}

\date{\today}

%---- Abstract  ----%
\begin{abstract}
\begin{description}
\item[Background:] 
The quenching of spin-isospin modes in nuclei is an important field of research in nuclear structure.
It has an impact on astrophysical reaction rates and on fundamental processes like neutrinoless double-$\beta$ decay.
Gamow-Teller (GT) and spinflip $M1$ strengths are quenched. 
Concerning the latter, the $J^\pi = 1^+$ resonance in the doubly magic nucleus $^{48}$Ca, dominated by a single transition, serves as a reference case.  
\item[Purpose:]
The aim of the present work is a search for weak $M1$ transitions in $^{48}$Ca with a high-resolution $(p,p^\prime)$ experiment at 295~MeV and forward angles including $0^\circ$ and a comparison to results from a similar study using backward-angle electron scattering at low momentum transfers in order to estimate their contribution to the total $B(M1)$ strength in $^{48}$Ca.
\item[Methods:] 
The spin-$M1$ cross sections of individual peaks in the spectra are deduced with a multipole decomposition analysis (MDA) and converted to reduced spin-$M1$ transition strengths using the unit cross section method.
For a comparison with electron scattering results,  corresponding reduced $B(M1)$ transition strengths are extracted following the approach outlined in J.~Birkhan {\it et al.}, Phys.~Rev.~C~{\bf 93}, 041302(R) (2016).  
\item[Results:] 
In total, 30 peaks containing a $M1$ contribution are found in the excitation energy region $7 - 13$~MeV.
The resulting $B(M1)$ strength distribution compares well to the electron scattering results considering different factors limiting the sensitivity in both experiments and the enhanced importance of mechanisms breaking the proportionality of nuclear cross sections and electromagnetic matrix elements for weak transitions  as studied here.    
The total strength of 1.14(7)~$\mu_{\rm N}^2$ deduced assuming a non-quenched isoscalar part of the $(p,p^\prime)$ cross sections agrees with the $(e,e^\prime)$ result of 1.21(13)~$\mu_{\rm N}^2$. 
A binwise analysis above 10~MeV provides an upper limit of 1.51(17)~$\mu_{\rm N}^2$.
\item[Conclusions:]
The present results confirm the previous electron scattering work that weak transitions contribute about 25\% to the total $B(M1)$ strength in $^{48}$Ca and the quenching factors of GT and spin-$M1$ strength are then comparable in $fp$-shell nuclei.
Thus, the role of meson-exchange currents (MECs) seems to be negligible in $^{48}$Ca, in contrast to $sd$-shell nuclei.
\end{description}
\end{abstract}

\maketitle

%---- Text  ----%
\section{Introduction}
\label{sec:Introduction}

Spinflip magnetic dipole excitations constitute an elementary excitation mode of nuclei and thus serve as an important test of nuclear structure models \cite{hey10}.
Knowledge of its properties is, e.g., important for modeling reaction cross sections in large-scale nucleosynthesis network calculations \cite{loe12} or neutral-current neutrino reactions in supernovae \cite{lan04,lan08}.  
Because the transitions mainly occur between spin-orbit partners they are also expected to show sensitivity to the evolution of single-particle properties leading to new shell closures in neutron-rich nuclei \cite{ots05,ots10}.

An investigation of the spinflip $M1$ strength also contributes to a resolution of the long-standing problem of quenching of the spin-isopin response in nuclei \cite{ost92}.
It represents the analog of the GT strength for $T_{\rm f} = T_{\rm i}$ (GT$_0$) transitions, where $T_{\rm i,f}$ denote the isospin of initial and final states, respectively. 
The same quenching mechanisms contribute to spinflip $M1$ and GT transitions but the magnitude can be different.
In light nuclei, meson-exchange currents (MEC) enhance the total $M1$ over the GT$_0$ strengths as demonstrated, e.g., for $N =Z$ nuclei in the $sd$-shell \cite{ric90,lue96,vnc97,hof02}.   
In $fp$-shell nuclei, comparable quenching factors for GT \cite{mar96} and $M1$ \cite{vnc98} transitions are needed in shell-model calculations to achieve agreement with the data.

Because of the particularly simple $[\nu 1f_{7/2}^{-1} 1f_{5/2}]$ particle-hole structure of $J^\pi = 1^+$ states, $M1$ strength in the doubly magic nucleus  $^{48}$Ca has been considered a reference case to study the quenching phenomenon \cite{hey10,tak88}.
The $M1$ strength is largely concentrated in a single transition to a state at 10.23~MeV.
It was first observed in inelastic electron scattering  \cite{ste80,ste83} with a reduced transition strength $B(M1)\myuparrow = 3.9(3)$~$\mu_{\rm N}^2$.
Recently, a much larger value $B(M1)\myuparrow = 6.8(5)$~$\mu_{\rm N}^2$ has been reported from a $^{48}$Ca$(\gamma,n)$ measurement at the HI$\gamma$S facility\cite{tom11} challenging our present understanding of quenching in microscopic models. 

The $J^\pi = 1^+$ states belonging to the spinflip $M1$ resonance in even-even nuclei can also be excited in small-angle inelastic proton scattering at energies of a few hundred MeV because angular momentum transfer $\Delta L = 0$ is favored in these kinematics and the spin-isospin part dominates over the isoscalar-spin and tensor parts of the proton-nucleus interaction \cite{lov81}.
The isoscalar giant monopole resonance populated through the dominant isoscalar interaction part resides at higher excitation energies and contributes little in the energy region where spinflip $M1$ transitions are expected.
Indeed, in pioneering experiments bumps were observed in forward-angle scattering spectra and identified as spinflip $M1$ resonance in heavy nuclei \cite{dja82,fre90}, but only recently high energy-resolution measurements at extreme forward angles including $0^\circ$ have become feasible \cite{tam09,nev11}.

At energies above 100~MeV,  a single-step reaction mechanism dominates in $(p,p')$ scattering in analogy to the $(p,n)$ and $(n,p)$ reactions \cite{ich06} implying a proportionality between the measured cross sections and the transition matrix elements.
This can be utilized to extract electromagnetic $M1$ transition strengths from such $(p,p')$ experiments based on isospin symmetry between the spinflip $M1$ mode and the GT mode excited in charge-exchange (CE) reactions \cite{bir16}. 
Using the data from Ref.~\cite{pol12} very good agreement with the $M1$ strength distribution in $^{208}$Pb extracted from electromagnetic probes \cite{las88} is obtained.
Application to the case of $^{48}$Ca resulted in an $M1$ transition strength compatible with the $(e,e^\prime)$ experiment and excluding the new $(\gamma,n)$ value.
 
For a quantitative interpretation of quenching in microscopic models the full $M1$ strength must be known experimentally. 
In $(e,e^\prime)$ scattering, 18 additional $M1$ transitions in $^{48}$Ca were identified \cite{ste83}.
Although individually weak ($\leq 0.15$~$\mu_{\rm N}^2$), they sum up to about 1.2~$\mu_{\rm N}^2$ which corresponds  to roughly 25\% of the total observed $B(M1)$ strength.
Most of these transitions were close to the detection limit of the $(e,e’)$ experiment, and there is considerable uncertainty about possible unobserved strength below the detection limit set by the radiative background and the high level density in the spectra at excitation energies above 10~MeV. 
The data used in the present work are not hampered by a large background.
We perform a multipole-decomposition analysis (MDA) \cite{pol12,kru15} of the $^{48}$Ca$(p,p’)$ data to extract the spinflip $M1$ cross sections, which can then be converted to $B(M1)$ transition strengths with the aid of the method described in Ref.~\cite{bir16}.
The result provides an independent constraint on the total $B(M1)$ strength in $^{48}$Ca.

The paper is organized as follows: Section \ref{sec:Data} gives a brief summary of the experiment, the data analysis and resulting spectra available for the MDA. 
Section \ref{subsec:MDAmethod} provides details of the MDA procedure, while Sec.~\ref{subsec:MDAresults} presents the corresponding results.
The method used to extract electromagnetic transition strengths from the spinflip $M1$ cross sections is described in Sec.~\ref{subsec:BM1method}.  
The electromagnetic $B(M1)\myuparrow$ strength distribution and its comparison with the $(e,e’)$ results is discussed in Secs.~\ref{subsec:M1strength} and \ref{subbsec:comparison}, respectively.  
Finally, conclusions are  given in Sec.~\ref{sec:conclusions}. 
\section{Experiment}
\label{sec:Data}

\subsection{Experimental details}
\label{sec:Experiment}

The $^{48}$Ca$(p,p')$ reaction was studied at the Research Center for Nuclear Physics in Osaka, Japan. 
A proton beam with currents 4 - 10\,nA was accelerated to an energy $E_\mathrm{p}=295$~MeV.
A self-supporting metallic $^{48}$Ca foil with an areal density of 1.87\,mg/cm$^2$ and an isotopic enrichment of 95.2\,\% served as target. 
Scattered protons were analyzed with the Grand Raiden magnetic spectrometer \cite{fuj99} placed under $0^\circ$, $2.5^\circ$, and $4.5^\circ$. 
Using dispersion-matching techniques an energy resolution of 25 keV (full width at half maximum) was achieved.
The experimental techniques of background suppression in $0^\circ$ scattering and the main steps for the raw-data analysis are described in Ref.~\cite{tam09}. 
Further details of the subtraction procedure and the analysis of the $^{48}$Ca data can be found in Ref.~\cite{bir15}.
 
\subsection{Spectra}
\label{sec:Spectra}

\begin{figure}[b]
        \centering
 \includegraphics[width=8.6cm]{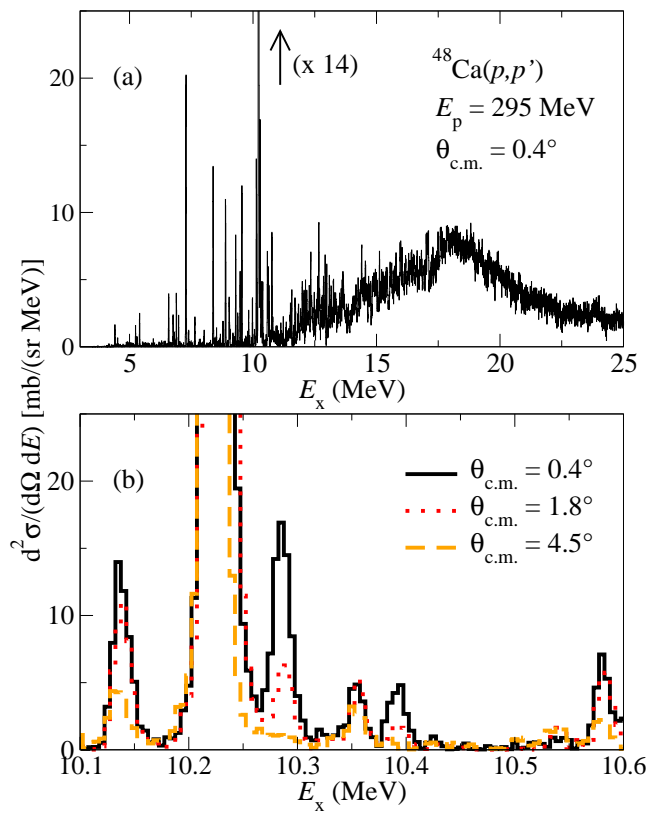}       
\caption{(color online).
(a) Spectrum of the $^{48}$Ca$(p,p')$ reaction at $E_0 = 295$~MeV and $\theta_{\rm c.m.} = 0.4^\circ$. 
The strong $M1$ transition at $E_\mathrm{x}=10.23$~MeV extends the scale by a factor of 14 and has a maximum cross section of about 350~mb/(sr MeV). 
(b) Spectra in the excitation energy range $E_{\rm x} = 10.1 - 10.6$~MeV for scattering angles $\theta_{\rm c.m.} = 0.4^\circ$ (solid line), $1.8^\circ$ (dotted line), and $4.5^\circ$ (dashed line).}
\label{fig:spectrum}
\end{figure}
The large acceptance of the Grand Raiden spectrometer permits a software decomposition of the data into spectra for up to three different angular bins for each spectrometer setting.
Thus, spectra of the double differential cross sections of the $^{48}$Ca$(p,p’)$ reaction are available at $\theta_{\rm c.m.} = 0.4^\circ,1.0^\circ, 1.8^\circ, 2.4^\circ, 3.3^\circ$, and $4.5^\circ$.
The target contained a non-negligible contribution from oxygen. 
It was subtracted from the spectra with the aid of $^{16}$O$(p,p^\prime)$ data measured in the same kinematics \cite{mat10} normalized to the well-known E2 transition in $^{16}$O at 6.917 MeV \cite{til93}.

Figure \ref{fig:spectrum}(a) shows the spectrum at $0.4^\circ$ as an example.
The by-far most strongly excited $J^\pi = 1^+$ state at 10.23~MeV is populated by a spinflip $M1$ transition.
Otherwise, at very forward angles relativistic Coulomb excitation of $J^\pi = 1^-$ states dominates the $(p,p')$ cross sections \cite{pol12,kru15,tam11,has15}.   
The resonance-like structure with a maximum at about 18.5~MeV is identified \cite{bir17} as the isovector electric giant dipole resonance consistent with data from a $^{48}$Ca$(e,e'n)$ experiment \cite{str00}.
Below 10~MeV the spectra are essentially free of instrumental background.
The stronger transitions visible in this energy region have all been observed in $(\gamma,\gamma’)$ experiments and identified to have dipole or quadrupole character \cite{har02,der14}.

An excerpt for the energy region $E_{\rm x} = 10.1 - 10.6$~MeV is presented in  Fig.~\ref{fig:spectrum}(b) with an overlay of spectra for different scattering angles.
Most of the observed peaks exhibit decreasing cross sections with increasing scattering angles $\theta_{\rm c.m.}$  characteristic for $E1$ or $M1$ transitions (note, however, the different behavior of the peak at 10.54~MeV).

\section{Multipole-decomposition analysis}
\label{sec:MDA}
\subsection{Method}
\label{subsec:MDAmethod}

In order to extract the cross-section part of the spectra due to $M1$ transitions, a  MDA has been performed.
In the MDA, the experimental angular distribution of the cross sections of a particular transition or an energy bin in the spectra are fitted to a sum of theoretical angular distributions for different possible multipolarities calculated in distorted-wave Born approximation (DWBA)
\begin{equation}
\left( \frac{d\sigma}{d\Omega}\right)_{\mathrm{th}} \left(\theta\right)=\sum_{E^\lambda/M^\lambda} a(E^\lambda/M^\lambda)\left( \frac{d\sigma}{d\Omega}\right) \left(\theta,E^\lambda/M^\lambda\right),
\label{eq:tho-distribution}
\end{equation}
in which $a(E^\lambda/M^\lambda) \geq 0$ denotes the weighting factors for each multipolarity. 
MDA is routinely applied in investigations of electric and spinflip giant resonances with hadronic reactions, see Refs.~\cite{ich06,har01} for examples. 

Calculations were performed with the code NLopt \citep{joh13} for all possible combinations but limited to one theoretical angular distribution for each multipolarity.
The program tries to minimize the checksum $S^2$ 
\begin{equation}
S^2=\sum_{i=1}^N\left( \frac{ \left( \frac{d\sigma}{d\Omega}\right)_\mathrm{th} \left(\theta_i\right) - \left( \frac{d\sigma}{d\Omega}\right)_\mathrm{exp} \left(\theta_i\right) }{u\left( \left( \frac{d\sigma}{d\Omega}\right)_\mathrm{exp} \left(\theta_i\right)\right)   }\right) ^2
\end{equation} 
weighted by the uncertainty $ u((d\sigma/d\Omega)_\mathrm{exp}(\theta_i))$ of the experimental cross sections.
Here, $N$ denotes the number of data points. 
An average over the $a(E^\lambda/M^\lambda)$ values for a given multipolarity is determined via
\begin{equation}
 \left\langle \frac{d\sigma}{d\Omega}\left(0^\circ,a(E^\lambda/M^\lambda) \right) \right\rangle   =\frac{\sum_lS^{-2}_{\mathrm{red,}l}\cdot  \frac{d\sigma}{d\Omega} \left(0^\circ, a(E^\lambda/M^\lambda) \right)_l }{\sum_lS^{-2}_{\mathrm{red,}l}},
\label{eq:wq-mittlung}
\end{equation}
where the reduced checksum $S_\mathrm{red}^2=S^2/(N-f)$ is introduced to compare results of multipole decompositions with a different number $f$ of allowed theoretical angular distributions. 

Theoretical angular distributions of the cross sections for different multipolarities were computed with the program code DWBA07 \cite{ray07} using wave functions from the quasiparticle phonon model (QPM) \cite{sol92} and the effective Love-Franey proton-nucleus interaction \cite{lov81}. 
It has been shown that the QPM provides a very good description of nuclear structure in heavy nuclei near shell closures (see, e.g., Refs.~\cite{pon99,rye02,sav08}) including the momentum transfer dependence of form factors in electron scattering and angular distributions in proton scattering \cite{pol12,bur07,wal11}.
For $^{48}$Ca, DWBA cross sections for the excitation of states with different spin and parity have been calculated in the  one-phonon approximation.
Taking the strongest excited states on the one-phonon level the angular dependence of the different modes is entirely governed by the transferred angular momentum. 

\begin{figure}[b]
        \centering
 \includegraphics[width=8.6cm]{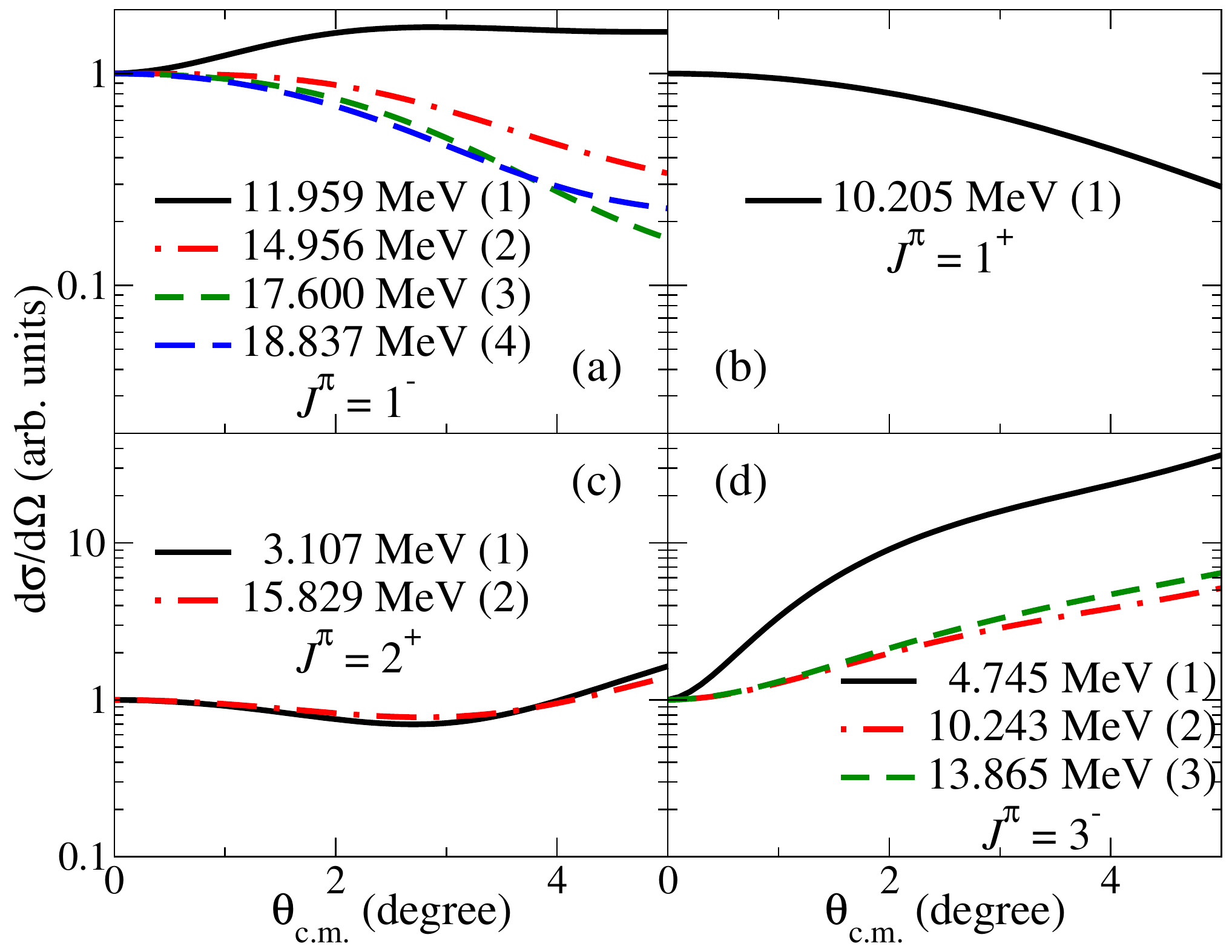}       
\caption{(color online).
Calculated angular distributions used in the MDA for  $E1$, $M1$, $E2$, and $E3$ transitions populating  (a) $J^\pi = 1^-$, (b) $1^+$,  (c) $2^+$, and (d) $3^-$ states in $^{48}$Ca, respectively, normalized at $\theta_\mathrm{c.m.}=0^\circ$.}
\label{fig:DWBA}
\end{figure}
Because of the small experimental momentum transfers only angular momenta $\Delta L \leq 3$ are considered in the MDA.
Figure \ref{fig:DWBA} summarizes the QPM results for $E1$, $M1$, $E2$, and $E3$ transitions populating $J^\pi = 1^-, 1^+, 2^+$, and $3^-$ states in $^{48}$Ca, respectively.
The DWBA calculations include Coulomb scattering, and the interference with nuclear scattering important for $E1$ transitions, where Coulomb excitation dominates, leads to a greater variety of possible $E1$ angular distributions.
In contrast, $M1$ transitions are described by a ‘universal’ curve in the small-$q$ range of the data independent of the particular nucleus. 
This is also approximately true for $E2$ excitations.
In the forward angle range studied here, spin-dipole $(\Delta L = 1, \Delta S = 1)$ transitions exhibit an angular dependence very similar to some of the theoretical curves for $E3$ transitions and therefore are not explicitly included in the fits.    

\begin{figure}[tbh]
      \centering
 \includegraphics[width=8.6cm]{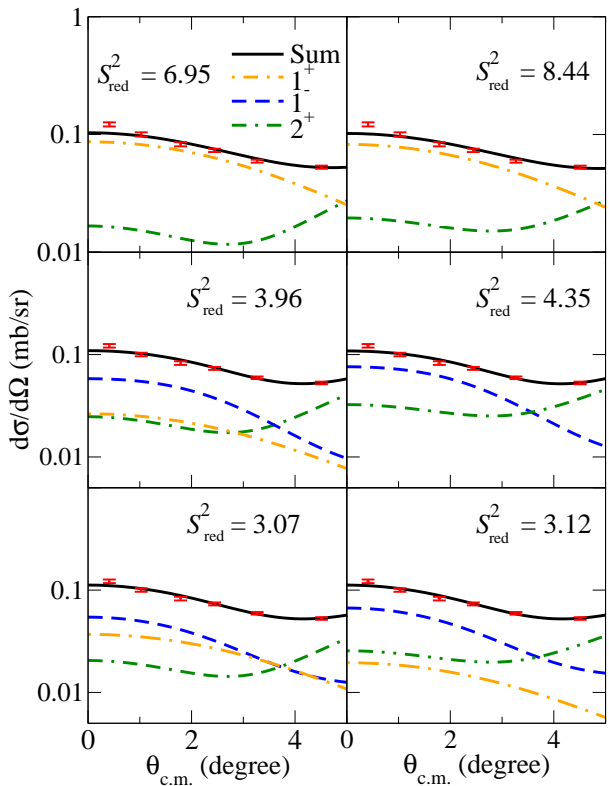}         
\caption{(color online).
Examples of the MDA for the peak at $E_\mathrm{x}=12.275$~MeV using different combinations of theoretical angular distributions from Fig.~\ref{fig:DWBA}.}
\label{fig:entfaltung12275}
\end{figure}

As an example, Fig.~\ref{fig:entfaltung12275} shows MDA results for different combinations of $E1$, $M1$, and $E2$ theoretical angular distributions for the transition to a state at 12.275~MeV. 
%The $E3$ contributions are not visible because the coefficients $a(E3) \approx 0$ in all fits.
The best $S^2_{\rm red}$ values are obtained for dominant $E1$ cross sections. 
However, smaller $M1$ and $E2$ contributions are also needed for optimum $S^2_{\rm red}$ values (bottom row).
The need for the latter stems from the slow fall-off of cross sections at the largest angles, which cannot be described by either $E1$ or $M1$ angular distributions. 
The $E1$ contributions are almost negligible in the fits shown in the top row which use model angular distribution (1) from Fig.~\ref{fig:DWBA}.
Here, the $M1$ component dominates but the overall fit is poorer.
The $M1$ cross section part at $0^\circ$ is determined from Eq.~(\ref{eq:wq-mittlung}) weighting with the $S_\mathrm{red}^2$ values.

\subsection{Results}
\label{subsec:MDAresults}

Two types of the MDA are discussed in the following, a single-peak analysis in the excitation energy range $E_{\rm x} = 7 -13$~MeV and a binwise analysis for excitation energies $10 - 13$~MeV.
The energy range was defined based on the following arguments: 
At lower $E_{\rm x}$ the number of excited states is small because of the double shell closure of $^{48}$Ca and one can assume that the spectroscopic information is sufficiently complete \cite{bur06}.
Shell-model calculations can provide a detailed description of the $M1$ strength distribution and predict a compact resonance concentrated in the investigated energy range \cite{vnc98}.  

\begin{figure}[tbh]
      \centering
 \includegraphics[width=8.6cm]{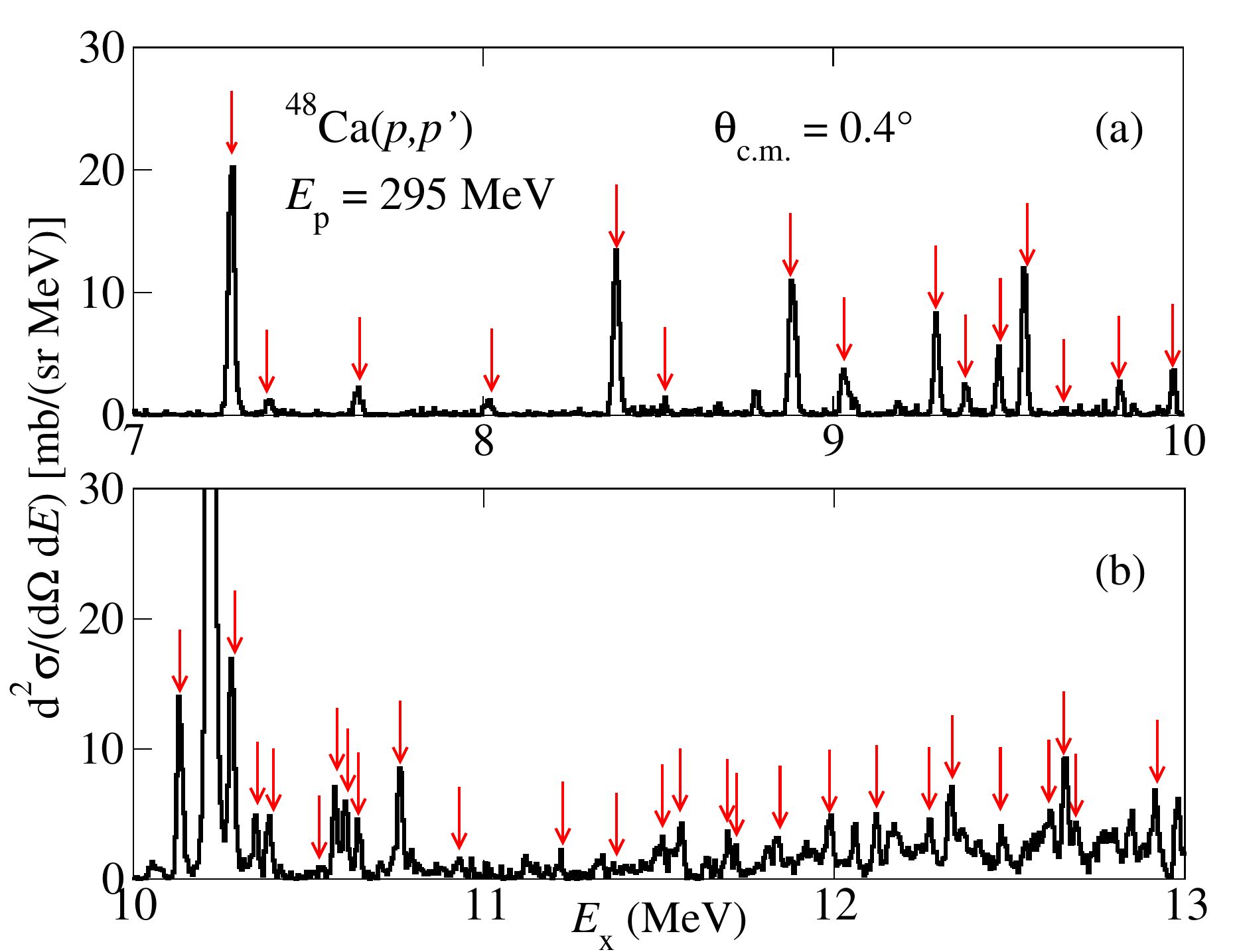} 
\caption{(color online).
Extended view of the spectrum of Fig.~\ref{fig:spectrum} between 7 and 13~MeV. 
The arrows indicate the energies of structures investigated in the single-peak analysis. }
\label{fig:Peak-Markierung}
\end{figure}
Figure \ref{fig:Peak-Markierung} presents an extended view of the spectrum shown in Fig.~\ref{fig:spectrum} for the energy region $7-13$~MeV.
In total, 41 structures indicated by red arrows are considered in the single-peak analysis. 
These have been identified in all 6 spectra.
(Note that results for the prominent transition at 10.23~MeV have been reported in Ref.~\cite{bir16} and therefore are not considered here).
Below 10~MeV, the spectrum is background-free and almost all peaks visible in the most-forward angle spectrum are included.
One exception is the peak at 8.8~MeV which was only observed in the spectrum shown.
In this energy region, available spectroscopic information \cite{bur06} is included as a guide of possible multipolarities.   
Above the neutron threshold ($S_n = 9.9953$~MeV), the level density and level widths increase such that the transitions are not always fully resolved.
Thus, combinations of all possible multipolarities are considered. 
Alternatively, a binwise analysis representing an upper limit of the possible $M1$ cross sections is performed.

\subsubsection{Single-peak analysis between 7 and 10~MeV}
\label{subsubsec:spa710}

The state density in the energy range from 7 to 10~MeV is small and the peaks are well separated. 
However, the available spectroscopic information \cite{bur06} indicates that within the energy resolution of the experiment and the systematic uncertainties of the energy calibration many peaks may correspond to doublets, even neglecting the possible excitation of states with $J > 3$.  
Accordingly, the MDA is performed assuming a single multipolarity or a combination of two multipolarities.  
Furthermore, data from the $^{48}$Ca$(\gamma,\gamma ')$ reaction \cite{har02}, which selectively excites dipole and to a lesser extent $E2$ transitions, and the $(e,e’)$ results \cite{ste83} are used as a guide for possible $E1$ and $M1$ transitions.
The electric character of all dipole transitions observed in Ref.~\cite{har02} has been shown in a subsequent experiment at Hi$\gamma$S using polarized photons \cite{der14}.

To check the possible correspondence of excitation energies $E_{\rm NDS}$ from the Nuclear data Sheets (NDS) \cite{bur06} with the $^{48}$Ca($p,p'$) results, the condition \cite{wei94}
\begin{equation}
\frac{\left| E_\mathrm{x}-E_\mathrm{NDS} \right|}{\sqrt{u^2(E_\mathrm{x})+u^2(E_\mathrm{NDS})}}\leq\sqrt{2}
\label{eq:vertraeglichkeitspruefung}
\end{equation}
is used.
Here, $u$ stands for the quoted uncertainties. 
The absolute accuracy of excitation energies in the $(p,p’)$ data is $\pm 10$ keV. 
A summary of the comparison and the most likely assignments is given in Table \ref{tabelle-Energievergleich}. 
\begin{table}
\centering
\caption{Comparison of states excited in the $^{48}$Ca$(p,p')$ reaction and candidates from the NDS \cite{bur06} fulfilling the condition of Eq.~(\ref{eq:vertraeglichkeitspruefung}).
The accuracy of excitation energies from the $(p,p^\prime)$ data is $\pm 10$ keV.}
\begin{tabular}{cccc}
\hline
\hline
Present & \multicolumn{2}{c}{Ref.~\cite{bur06}} & MDA \\ 
$E_\mathrm{x}$ & $E_\mathrm{x}$  & $J^\pi$ &   \\
(MeV) & (MeV) & & \\
\hline
\multirow{2}{*}{7.285}  & 7.296 &       ($2^+$) & \multirow{2}{*}{no fit} \\
      & 7.299 &     $1^{(-)}$ &  \\
      \hline
7.385  &  7.371    & ($\leq 2$)   & $3^-$(+$1^-$) \\
\hline
\multirow{2}{*}{7.648}  & 7.652        &  $3^-$ & \multirow{2}{*}{$3^-$(+$1^+$)} \\
      & 7.656     &  $1$  &  \\
      \hline
8.018  & 8.028    &  $2^+$ & ($2^+$) \\
\hline
\multirow{2}{*}{8.385}  & 8.385    & $1^-$ &  \multirow{2}{*}{$1^-$+$3^-$}  \\
      & 8.386     &  $3^-$ &  \\
      \hline
\multirow{2}{*}{8.520}  & 8.518      &  $1,2$  & \multirow{2}{*}{$3^-$+$1^+$}  \\
      & 8.522      &  $3^-$ &  \\
      \hline
8.893  & 8.883     &  $2^+$ & ($2^+$)  \\
      \hline
\multirow{2}{*}{9.043}  & 9.034      &  $1^-$ & \multirow{2}{*}{$1^-$+$2^+$} \\
      & 9.049      &  $2^+$ &  \\
      \hline
\multirow{2}{*}{9.298}  & 9.292      &  $1^-$ & \multirow{2}{*}{$1^-$+$2^+$}  \\
      & 9.295      &  $2^+$ &  \\
      \hline
9.383  & 9.392       &  $(1^+,2^+)$ & $1^+$+$2^+$ \\
\hline
9.475  & 9.473      &  $1^-$ & $1^-$+$3^-$  \\
\hline
\multirow{2}{*}{9.548}  & 9.546      &  $1^-$ & \multirow{2}{*}{$1^-$+$3^-$} \\
      & 9.550   &  $(3^-)$ &  \\
      \hline
9.653  & 9.638       &  $2^-,3^-,4^-$ & no fit \\
\hline
9.823  & 9.810        &  $(1)^-$ & no fit \\
\hline
9.973  & 9.954       & $1^+$  & $1^+$  \\
\hline
\hline
\end{tabular}
\label{tabelle-Energievergleich}
\end{table}
\begin{figure}[b]
        \centering
 \includegraphics[width=8.6cm]{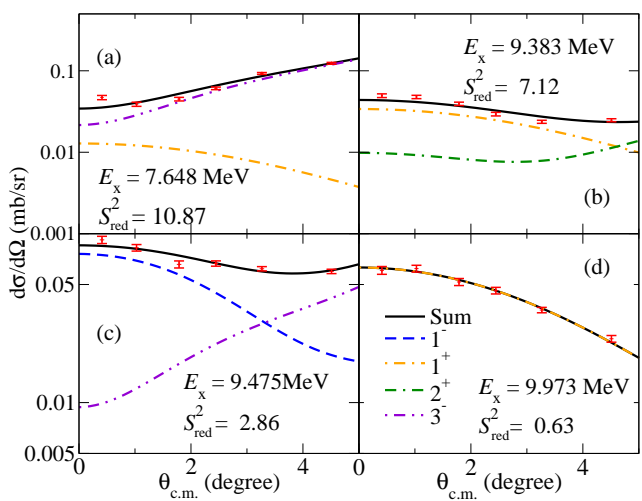}     
\caption{(color online).
Examples of the MDA with the smallest $S^2_{\rm red}$ values for the transitions to the states at (a) 7.648 MeV, (b) 9.383 MeV, (c) 9.475 MeV, and (d) 9.973 MeV.
}
\label{fig:entfaltung_8520_9823}
\end{figure}
Examples of the MDA results are presented in Fig.~\ref{fig:entfaltung_8520_9823} showing the fit with the smallest $S^2_{\rm red}$ value for each case. 
The transition to the peak at 7.648~MeV exhibits an angular distribution increasing with scattering angle suggesting $L > 1$.
The best description is obtained assuming $E3$.
The most forward angles indicate the presence of dipole strength but because of the dominance of the higher-$L$ component the $S^2$ values do not distinguish between $E1$ and $M1$. 
The angular distribution of the peak at 9.383~MeV is consistent with a $M1$ character except for the largest angle, which requires inclusion of an $E2$ part. 
The most forward angles confirm the $E1$ assignment for the transition to the state at 9.475~MeV from the $(\gamma,\gamma')$ data but an additional $E3$ component is needed for a reasonable fit at larger angles.
Finally, an excellent fit is obtained assuming $M1$ for the excitation of the state at 9.973~MeV.  

In the following, the results for each peak are discussed briefly. 

\noindent
{\it 7.285 MeV.} 
There is no combination of any two multipolarities which would provide a good fit to the data and hence no $M1$ component is considered. 
The $^{48}$Ca$(\gamma,\gamma ')$ experiments find a $1^-$ state a 7.299~MeV and Ref.~\cite{bur06} quotes a state at 7.296~MeV with $J^\pi = (2^+)$ but the correspondence is uncertain considering the energy difference.
  
\noindent
{\it 7.385 MeV.} 
The multipole decomposition gives the best fit for a $3^-$ state or a mixture of $3^-$ and $1^-$. 
There is no indication for an $M1$ part from the fit.
A pure ground-state transition from a state at 7.371 MeV was identified in the $(n,n^\prime \gamma$) reaction \cite{van92}, thus $J \leq 2$.
The comparison with the fit suggests $ J^\pi = 1^-$, however, no transition around this energy was observed in the ($\gamma,\gamma^\prime$)  data \cite{har02}.  

\noindent
{\it 7.648 MeV.}
Two states with $J^\pi = 3^-$ and $J = 1$ are known within the experimental energy uncertainty of the peak. 
The MDA favors fits of the combinations ($E3$, $E1$) and ($E3$, $M1$) with comparable $S^2_{\rm red}$ values, the latter shown in the top left of Fig.~\ref{fig:entfaltung_8520_9823}. 
As mentioned above an E3 contribution is necessary to describe the rising of the angular distribution for larger angles. 
Assuming $M1$ for the dipole part the averaged cross section at $0^\circ$ is 0.015(9)~mb/sr. 

\noindent
{\it 8.018 MeV.}
The angular distribution is described best if the multipole decomposition contains an $E2$ part. 
This agrees with the assignment of Ref.~\cite{bur06}.
However, a description of the data in terms of a pure $E2$ transition is rather poor.

\noindent
{\it 8.385 MeV.}
The $^{48}$Ca$(\gamma,\gamma ')$ experiment found the excitation of a $1^-$ state at $E_\mathrm{x}=8.385$~MeV, and  a $3^-$ state at 8.386~MeV is quoted in Ref.~\cite{bur06}. 
Indeed, the combination of $E1$ and $E3$ excitations provides one of the lowest $S^2_{\rm red}$ values confirming these assignments.

\noindent
{\it 8.520 MeV.}
The $^{48}$Ca$(\gamma,\gamma ')$ data show population of a state at \mbox{$E_\mathrm{x}=8.518$~MeV} but canot decide on the spin ($J = 1$ or 2).
Excitation of a $3^-$ state at 8.522 MeV has been observed in many reactions \cite{bur06}.
The combination of dipole-plus electric octupole gives the best fits in the MDA but no $E1$/$M1$ distinction is possible.  
Assuming $M1$ character for the dipole part a cross section of  $0.012(5)$~mb/sr at $0^\circ$ is extracted.  

\noindent
{\it 8.893 MeV.}
The $(\gamma,\gamma ')$ measurements find a $2^+$ state at $E_\mathrm{x}=8.883$~MeV consistent with results from other reactions \cite{bur06}.
$E1/E2$ and $M1/E2$ combinations provide a superior fit to the assumption of a pure $E2$ transition.
However, both the corresponding $B(E1)$ or $B(M1)$ (deduced with the method described below) value should have led to a signal visible in the $(\gamma,\gamma ')$ data. 
Thus, no possible $M1$ component of the cross sections is considered.
 
\noindent
{\it 9.043 MeV.}
The available spectroscopic information suggests a combined excitation of a $1^-$ state at 9.034~MeV and a $2^+$ state at 9.049~MeV.
Indeed, an $E1$/$E2$ combination leads to the best fit of the $(p,p^\prime)$ data. 
 
\noindent
{\it 9.298 MeV.}
Previous data suggest an excitation of close-lying $1^-$ and $2^+$ states \cite{bur06}. 
This is consistent with the MDA favoring an $E1$/$E2$ combination although $E1$/$E3$ gives a similar $S^2_{\rm red}$.
In any case, there is no indication of a $M1$ contribution.

\noindent
{\it 9.383 MeV.}
There is an indication of a corresponding transition in the backward-angle $(e,e^\prime)$ data \cite{ste83}.
A fit assuming a $M1$ transition leads to $S^2_\mathrm{red}=16.1$ but the fit is improved allowing for $M1$ plus $E2$.
The best fit ($S^2_{\rm red} = 0.60$) is obtained with an $E1$/$E2$ combination.
However, we exclude an $E1$ contribution because no corresponding peak was seen in the $(\gamma,\gamma ')$ data.
When the partial $E1$ cross section from the MDA is converted to a $B(E1)$ value normalizing to the theoretical value from the QPM for the corresponding angular distribution, one ends up with a transition strength well above the sensitivity limits of the $(\gamma,\gamma ')$  experiments \cite{har02,der14}.
This is not the case assuming a dominant $M1$ transition.  
The corresponding $M1$ cross section at $0^\circ$ amounts to $0.035(1)$~mb/sr.

\noindent
{\it 9.475 MeV.}
The excitation energy is consistent with observation of a transition at 9.473~MeV in the $(\gamma,\gamma ')$ experiments suggesting a pure $E1$ character.
Again the $(p,p^\prime)$ data at larger angles require inclusion of a $L > 1$ multipolarity (cf.\ bottom left of Fig.~\ref{fig:entfaltung_8520_9823}).
Fits with a $M1$ instead of an $E1$ part lead to larger $S^2_{\rm red}$ values.

\noindent
{\it 9.548 MeV.}
The spectroscopic information suggests again simultaneous excitation of $1^-$ (9.546~MeV) and $3^-$ (9.550~MeV) states consistent with the MDA results. 

\noindent
{\it 9.653 MeV} and {\it 9.823 MeV.}
There is no prior information from other experiments \cite{bur06} which could be related to these small peaks.
The MDA does not allow for unique assignments.
Thus, no possible $M1$ contributions are considered.

\noindent
{\it 9.973 MeV.}
The transition to the state at 9.973~MeV shows clear $M1$ character as demonstrated in the bottom right part of  Fig.~\ref{fig:entfaltung_8520_9823}.
An $E1$ character is also excluded by the absence of a corresponding transition in the $(\gamma,\gamma^\prime)$ data.
The $^{48}$Ca(e,e') experiment finds the excitation of a $1^+$ state at 9.954~MeV \cite{ste83}.
Despite about 20 keV difference of the centroid energies one may argue that both experiments may have seen the same $1^+$ state considering the systematic uncertainties of the respective energy calibrations.   
The $M1$ cross section at $0^\circ$ is 0.064~mb/sr.

\subsubsection{Single-peak analysis between 10 and 13~MeV}
\label{subsubsec:spa1013}

Above 10\,MeV the increase of the level density makes an interpretation of the peaks as an excitation of a single state unlikely.  
Guided by the dominance of $E1$ and $M1$ cross sections in comparable $(p,p’)$ data for heavier nuclei \cite{pol12,kru15,tam11,has15}, combinations of $E1$/$M1$/$E2$ and $E1$/$M1$/$E3$ transitions are considered. 
The fits are constrained to at most three theoretical angular distributions because of the limited number of data points.  
An example of the procedure is illustrated in Fig.~\ref{fig:entfaltung12275} for excitation of the 12.275 MeV peak.
A summary of the partial $M1$ cross sections at $0^\circ$ deduced from the MDA is given in Table \ref{tabelle-Vergleich-ee}.

\begin{figure}[tbh]
       \centering
 \includegraphics[width=8.6cm]{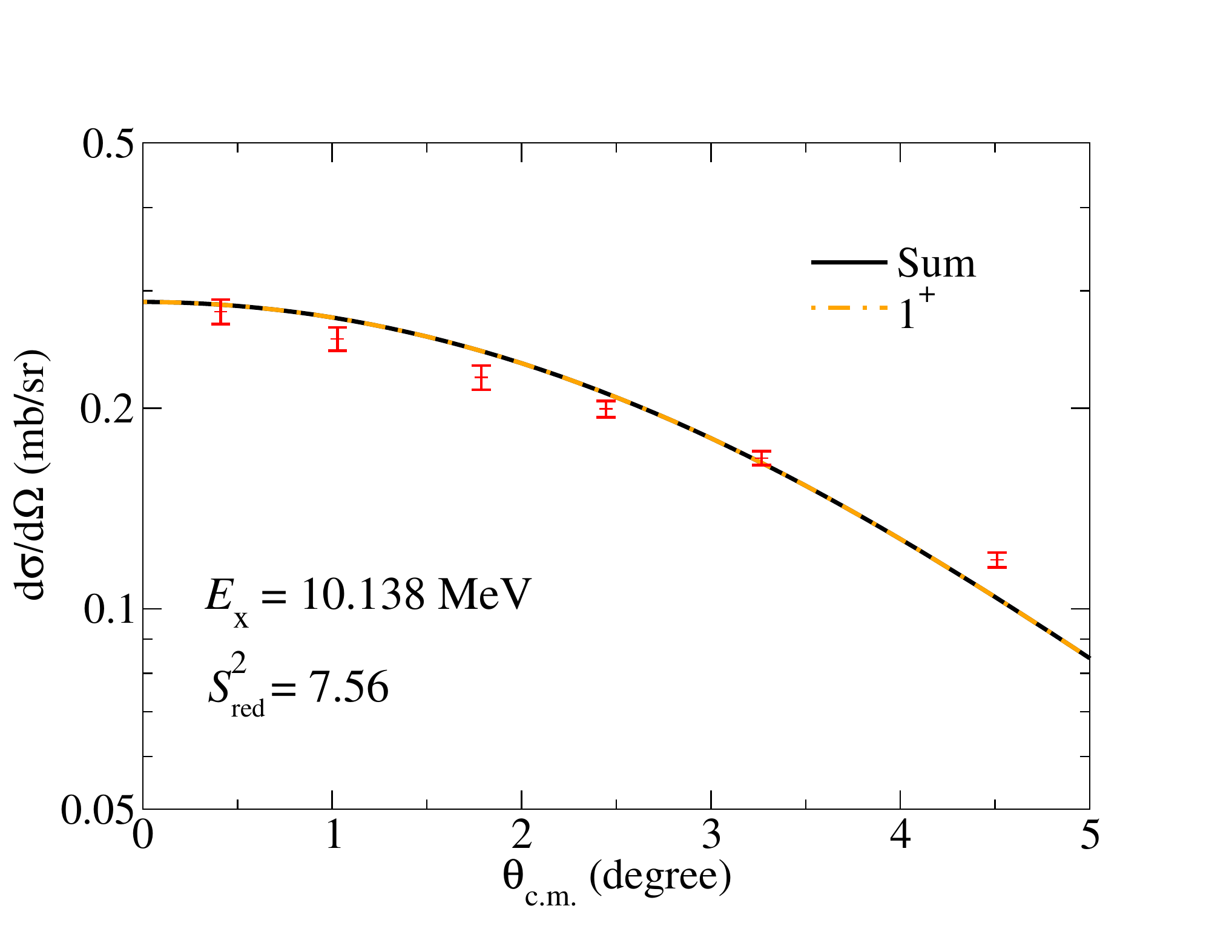}          
\caption{(color online).
Angular distribution of the transition to the state at 10.138~MeV.
The MDA clearly favors a pure $M1$ character.}
\label{fig:entfaltung_9383_9475_9973}
\end{figure}
The excitation of the state at 10.138~MeV is an exception, where the data and the good correspondence with the electron scattering result suggests a pure $M1$ transition.
The corresponding experimental angular distribution and the fit are depicted in Fig.~\ref{fig:entfaltung_9383_9475_9973}.

\subsubsection{Binwise analysis between 10 and 13~MeV}
\label{subsubsec:bwa}

Above the neutron threshold a small physical background, most likely due to quasifree reactions, is observed in the data. 
Together with the high level density this may lead to a situation where weak transitions can no longer be resolved. 
In order to estimate possible missing $M1$ cross section parts, a binwise MDA of the total cross sections is performed. 
The background component was not modeled separately but it was assumed that the combination of different multipoles allowed in the MDA can mimic its angular dependence.  
 
The prominent excitation of the state at 10.23~MeV is again excluded.
The binwise analysis covers the energy range $10.26 - 13$~MeV divided into 10 bins.

\section{$B(M1)$ strength distribution in $^{48}{\rm Ca}$}
\label{sec:Extraction of the B(M1) strength}

In this section the isovector spinflip-$M1$ (IVSM1) transition strength distribution is extracted from the $M1$ cross sections at $0^\circ$. 
Using approximations explained below the corresponding electromagnetic $B(M1)\myuparrow$ strength distribution in $^{48}$Ca can be derived.
Only a brief summary of the method is given here since it has been presented in Refs.~\cite{bir15,bir16} including a discussion of the underlying approximations and estimates of the systematic uncertainty.
%In all of the following, the prominent $M1$ transition to the state at 10.23~MeV is again left out since it has already been discussed in Ref.~\cite{bir16}. 

\subsection{Extraction of $B(M1)$ strength from the $(p,p^\prime)$ data}
\label{subsec:BM1method}

For incident energies high enough to ensure the dominance of one-step reactions one can relate the proton inelastic scattering cross sections at $0^\circ$ to the IVSM1 strength 
\begin{equation}
\frac{\mathrm{d}\sigma}{\mathrm{d}\Omega}(0^\circ) = \hat{\sigma}_{M1} F(q,E_{\rm x}) B(M1_{\sigma \tau}), 
\label{eqm1} 
\end{equation}
where $\hat{\sigma}_{M1}$ is a nuclear-mass dependent factor (the so-called unit cross section), $F(q,E_{\rm x})$ a kinematical factor correcting for non-zero momentum and energy transfer, and  $B(M1_{\sigma\tau})$ denotes the reduced IVSM1 transition strength.
%Obviously, the kinematical factors in Eqs.~(\ref{eqgt}) and (\ref{eqm1}) differ for isobaric analog states.
The kinematical correction factor is determined by DWBA calculations and the extrapolation of the $M1$ cross section part at finite angles deduced with the MDA to $0^\circ$ is achieved with the aid of the theoretical $M1$ angular distribution shown in Fig.~\ref{fig:DWBA}. 

The unit cross section is taken from a corresponding relation for analog Gamow-Teller (GT) strengths in $(p,n)$ charge-exchange reactions \cite{tad87,zeg07}.
At the very small momentum transfers considered here, isospin symmetry \cite{fuj11} suggests $\hat{\sigma}_{\mathrm{M1}} \simeq \hat{\sigma}_\mathrm{GT}$.
The systematics of $\hat{\sigma}_\mathrm{GT}$ for the $(p,n)$ reaction at incident energies comparable to the present experiment has been studied in Ref.~\cite{sas09}. 
A simple mass-dependent parameterization is given there, which allows to extract $\hat{\sigma}_{\mathrm{M1}}$ for $^{48}$Ca.
The resulting value is consistent with a recent analysis of its mass dependence in lighter nuclei \cite{mat15} extrapolated to mass number 48. 

As discussed in Ref.~\cite{bir16}, several effects can break the proportionality between cross section and matrix element in  Eq.~(\ref{eqm1}).
While most of these are either small or taken into account in the MDA, a general problem are coherent $\Delta L = 2$ contributions to the excitation of $1^+$ states invoked by the tensor part of the interaction.
This problem has been investigated in Ref.~\cite{zeg06}  for GT transitions in the framework of a shell-model study.
The lowest $B(M1_{\sigma\tau})$ strengths found in the present work correspond to GT strengths of the order of 0.001.
This implies systematic uncertainties for individual transitions of about 30\%, maybe up to 50\% for the weakest strengths.
However, since the interference has random sign \cite{zeg06,fay97} the effect on the total strength will be smaller.  
 
The corresponding electromagnetic $B(M1)\myuparrow$ transition strength
\begin{equation}
B(M1)\uparrow = \frac{3}{4\pi} \left | \langle f|| g_l^{\mathrm{IS}}\vec{l} + \frac{g_s^{\mathrm{IS}}}{2}\vec{\sigma} - (g_l^{\mathrm{IV}} \vec{l} + \frac{g_s^{\mathrm{IV}}}{2}\vec{\sigma})\tau_0||i \rangle \right |^2 \! \! \mu_\mathrm{N}^2
\label{eqbm1em}
\end{equation}
contains spin and orbital contributions for the isoscalar (IS) and isovector (IV) parts.
In the present work it is assumed that $B(M1)$ and $B(M1_{\sigma \tau})$ strengths are approximately the same based on the following arguments. 
Orbital $M1$ strength is connected to ground-state deformation \cite{hey10} and thus expected to be weak for the present case of a doubly magic nucleus. 
Because $g_s^{\rm{IV}} \gg g_s^{\rm{IS}}$ the isoscalar part is usually neglected.
Then, an analog electromagnetic transition strength $B^{\rm em}(\mathrm{M1}_{\sigma \tau})$ can be extracted from  the $(p,p^\prime)$ data 
\begin{equation}
B^{\rm em}(\mathrm{M1}_{\sigma \tau}) = \frac{3}{4\pi}\left(g_s^{\mathrm{IV}} \right)^2 B(\mathrm{M1}_{\sigma \tau}) \cong B(M1)
\label{eqbm1pp}
\end{equation}
and compared to $B(M1)$ strengths from electromagnetic probes.
Equation (\ref{eqbm1pp}) has, e.g., been successfully applied in the comparison of $M1$ strengths from electromagnetic and hadronic reactions in self-conjugate $sd$-shell nuclei \cite{ric90,lue96,vnc97}.

However, the strong transition in $^{48}$Ca has pure neutron character \cite{deh84} and it is assumed that this also holds for the weak transitions investigated here.
This assumption is motivated by a picture where the strong transition at 10.23 MeV acts as a doorway \cite{bbb98} and the fragmentation of the $M1$ strength distribution results from mixing with nearby complex (multi particle-multi hole) $1^+$ states. 
In such a scenario, the excitation probability is still determined by the amplitude of the doorway-state wave function.   

In the particular case of a pure neutron transition, the $\vec{\sigma}$ term in the electromagnetic operator, Eq.~(\ref{eqbm1em}), needs to be considered for the determination of the $B(M1)$ value because of the interference term.   
The IS contribution to the $M1$ $(p,p^\prime)$ cross sections of the 10.23~MeV transition amounts to 5.2(2.5)\% determined by a fit of theoretical angular distributions for IS and IV $1f_{7/2} \rightarrow 1f_{5/2}$ transitions \cite{bir16}.
The result is adopted for the present analysis.

Extraction of the analog electromagnetic strength requires the inclusion of quenching implemented through effective  $g$ factors $g_{s,\mathrm{eff}}^{\rm IS/IV} = q^{\rm IS/IV} \times g_{s}^{\rm IS/IV}$ in Eq.~(\ref{eqbm1em}), where $q$ denotes the magnitude of quenching. 
For $fp$-shell nuclei $q^{\rm IV} = 0.75(2)$ was determined in Ref.~\cite{vnc98}. 
A recent study indicates $g^{\mathrm{IS}}_{s,\mathrm{eff}} = g^{\mathrm{IS}}_s$ for the isoscalar spinflip $M1$ strength   in a series of $sd$-shell nuclei \cite{mat15}.
All results in the next section are derived with these quenching factors.
However, the isoscalar quenching factor may have a mass dependence.
As an extreme, one may assume $q^{\rm IV} = q^{\rm IS}$.
Then all $B(M1)$ strengths in Table~\ref{tabelle-Vergleich-ee} and in the binwise analysis would be larger by a factor 1.21.   

\subsection{Results}
\label{subsec:M1strength} 

\begin{figure}[b]
        \centering
  \includegraphics[width=8.6cm]{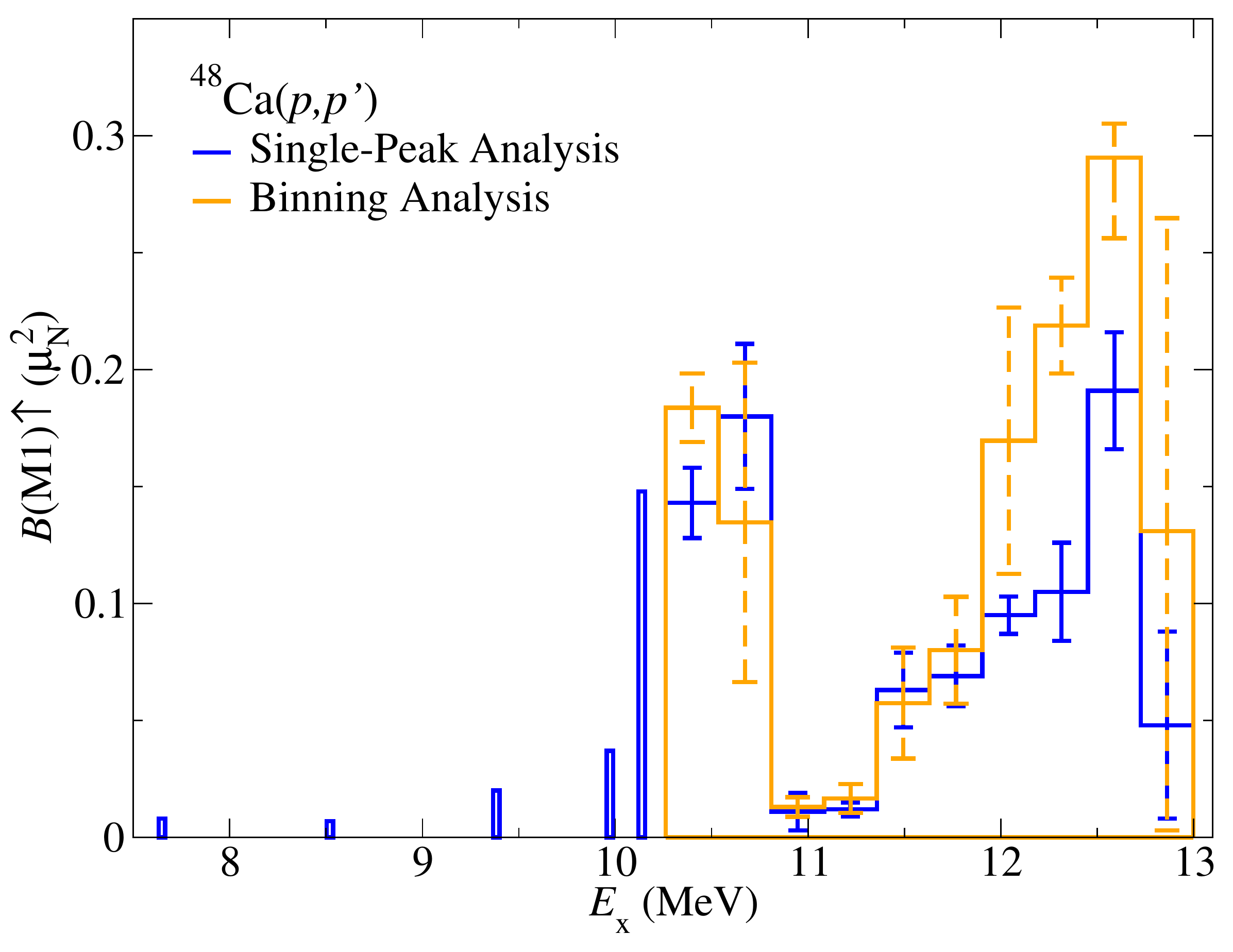}              
\caption{(color online).
$B(M1)\myuparrow$ strength extracted from the $^{48}$Ca$(p,p^\prime)$ data. 
Orange (light grey) histogram: binwise analysis above the prominent peak at 10.23~MeV. 
Blue (dark grey) histogram: single-peak analysis (Table~\ref{tabelle-Vergleich-ee}) summed in the same energy bins.}
\label{fig:Uebergangstarke}
\end{figure}
The $B(M1)\myuparrow$ strength distribution between 7 and 13~MeV deduced with the method explained in the previous subsection is displayed in Fig.~\ref{fig:Uebergangstarke}. 
The $B(M1)$ strength in the single-peak analysis is concentrated in the energy regions $10 - 11$~MeV and $12 - 13$~MeV, while strengths below 10~MeV are weak.
Table~\ref{tabelle-Vergleich-ee} summarizes the results.
The total strength amounts to  $B(M1)\myuparrow=1.14(7)$~$\mu_\mathrm{N}^2$.
A comparison with the binwise analysis shows agreement within error bars between 10 and 12~MeV but significantly larger strengths (up to a factor of two) of the latter at higher excitation energies.
The total strength of the binwise analysis including the single-peak analysis results at lower excitation energies is $B(M1)\myuparrow=1.51(17)$~$\mu_\mathrm{N}^2$ .

\begingroup
\squeezetable
\begin{table}[t]
\centering
\caption{Columns 1-3: Excitation energies (uncertainties $\pm10$~keV), cross sections at $0^\circ$ (uncertainties from MDA only), and $B(M1)\myuparrow$ strengths from the single-peak analysis of the $^{48}$Ca$(p,p^\prime)$ experiment.   
Columns 4-5: Excitation energies (uncertainties $\pm15$~keV) and $B(M1)\myuparrow$ strengths from the $^{48}$Ca$(e,e^\prime)$ experiment \citep{ste83}.  
Column 6: Ratio $R$ of $B(M1)$ strengths from electron and proton scattering for transitions which fulfill Eq.~(\ref{eq:vertraeglichkeitspruefung}).
}
\begin{tabular}{cccccc}
\hline
\hline
\multicolumn{3}{c}{Present work} & \multicolumn{2}{c}{Ref.~\cite{ste83}} & $R$ \\
$E_\mathrm{x}$ & $\sigma_{M1}(0^\circ)$ & $B(M1)\myuparrow$ & $E_\mathrm{x}$ & $B(M1)\myuparrow$ &   \\
MeV   & (mb/sr) & $\mu_\mathrm{N}^2$ & MeV   &  $\mu_\mathrm{N}^2$ & \\
\hline
7.648 & 0.015(9) & 0.008(5) &       &       &  \\
      &     &  & 7.696 & $<0.05$ &  \\
      &      & & 8.150 & $<0.05$ &  \\
8.520 & 0.012(5) & 0.007(3) &       &       &  \\
9.383 & 0.035(1) & 0.020(2) & 9.392 & $<0.07$ &  \\
      &     &   & 9.885 & $<0.09$ &  \\
9.973 & 0.063(0) & 0.037(3) & 9.954 & $<0.10$ &  \\
10.138 & 0.255(9) & 0.148(13) & 10.138 & 0.12(3) & 0.8(2) \\
10.288 & 0.137(1) & 0.080(8) &       &       &  \\
      &     &  & 10.330 & 0.09(4) &  \\
10.350 &0.069(22) & 0.040(13) & 10.354 & 0.08(4) & 2.0(1.2) \\
10.390 & 0.040(1) & 0.023(2) &       &       &  \\
10.538 & 0.017(4) & 0.010(3) &       &       &  \\
10.578 & 0.103(12) & 0.060(8) &       &       &  \\
10.610 & 0.053(6) & 0.031(4) &       &       &  \\
10.645 & 0.034(6) & 0.020(4) &       &       &  \\
10.763 & 0.102(48) & 0.059(29) & 10.782 & 0.12(4) & 2.0(1.2) \\
10.933 & 0.018(13) & 0.011(8) & 10.930 & 0.05(2) & 4.7(3.9) \\
11.225 & 0.020(5) & 0.012(3) &       &       &  \\
11.383 & 0.005(3) & 0.003(2) &       &       &  \\
      &     &  & 11.410 & $<0.09$ &  \\
11.513 & 0.036(26) & 0.021(15) & 11.490 & 0.15(3) & 7.2(5.4) \\
11.563 & 0.066(7) & 0.039(5) &       &       &  \\
11.695 &  0.043(15) & 0.025(9) &       &       &  \\
11.725 & 0.024(14) & 0.014(9) & 11.728 & 0.12(4) & 8.5(5.9) \\
11.843 & 0.051(6) & 0.030(4) &       &       &  \\
11.990 &0.079(5) & 0.047(5) &       &       &  \\
      &     &  & 12.055 & 0.08(3) &  \\
12.120 & 0.082(8) & 0.048(6) &       &       &  \\
12.275 & 0.059(32) & 0.035(19) & 12.270 & 0.10(5) & 2.8(2.1) \\
12.338 & 0.117(13) & 0.070(9) &  12.310     &   0.11(3)     &  1.4(0.4) \\
12.480 & 0.043(22) & 0.025(13) & 12.493 & 0.09(4) & 3.5(2.4) \\
12.623 & 0.090(32) & 0.054(20) &    &      &   \\
12.660 & 0.129(1) & 0.077(6) &       &       &  \\
12.693 & 0.059(7) & 0.035(5) & 12.700 & 0.10(5) & 2.8(1.5) \\
12.918 &  0.080(66) &   0.048(40)   & &       &  \\
\hline
\hline
\end{tabular}
\label{tabelle-Vergleich-ee}
\end{table}
\endgroup

\subsection{Comparison with the $(e,e^\prime)$ results}
\label{subbsec:comparison}  

\begin{figure}[t]
        \centering
  \includegraphics[width=8.6cm]{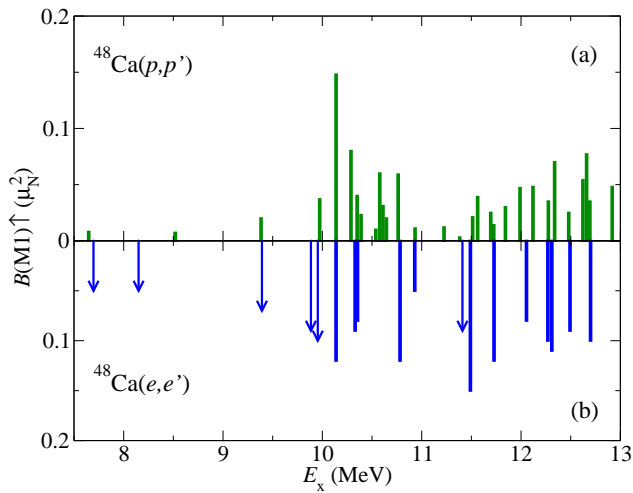}           
\caption{(color online).
Comparison between the $B(M1)$ strength distributions in $^{48}$Ca deduced from (a) the $(p,p')$ reaction (present work) and (b) the $(e,e')$ reaction \cite{ste83}.
Arrows indicate upper limits.  
The prominent transition to the state at $E_\mathrm{x}=10.23$~MeV is excluded.}
\label{fig:comparison}
\end{figure}

A comparison of the $B(M1)$ strength distribution deduced from the single-peak analysis with the results from the $^{48}$Ca$(e,e')$ experiment \citep{ste83} is presented in Fig.~\ref{fig:comparison} and Table \ref{tabelle-Vergleich-ee}.
Leaving out transitions from Ref.~\cite{ste83}, for which only upper limits are given, we find correspondence with all but two transitions identified in the $(e,e')$ data based on the criterion Eq.~(\ref{eq:vertraeglichkeitspruefung}).
Exceptions are the transitions to states at 10.330~MeV and 12.055~MeV in Ref.~\cite{ste83}.
The former fulfills Eq.~(\ref{eq:vertraeglichkeitspruefung}) when assigned to the peak observed in proton scattering at 10.350~MeV. However, an assignment to the 10.354~MeV transition seen in electron scattering is considered more likley.  

The strengths from electron scattering tend to be larger (see the ratio $R$ of electron-to-proton scattering strengths in Table \ref{tabelle-Vergleich-ee}) but are still consistent within error bars in many cases.
This is particularly true if one relaxes condition (\ref{eq:vertraeglichkeitspruefung}) somewhat and, e.g., relates the strength of the transition seen at 12.700~MeV in electron scattering to the sum of the transitions at 12.660 and 12.693~MeV in proton scattering. 
Possible differences between the strengths may be related to the assumptions underlying the analysis of the $(p,p^\prime)$ data explained in Section~\ref{subsec:BM1method}. 
Some of these could also affect the average ratio of $R$.
For example, orbital contributions -- although shown to be weak \cite{ric85} -- could lead to a systematic enhancement of the $B(M1)$ strength by constructive interference with the spin part, since the dominant shell-model configurations are the same in all $1^+$ states.
For the same reason one can also speculate about a systematic reduction of $B(M1_{\sigma \tau})$ due to the interference of $\Delta L =2$ contributions discussed above.     
While the shell-model study of $^{26}$Mg showed a random sign of the mixing in an open-shell nucleus \cite{zeg06}, this may be different in a case, where the wave functions of all excited $1^+$ states are similar.   

The present analysis finds 30 $M1$ transitions compared to 18 seen in Ref.~\cite{ste83}.
This may be related to the different sensitivity thresholds in both experiments.
For the $(e,e')$ data a statistical limit due to the radiative tail in the spectra and difficulties to distinguish  $M1$ and $M2$ form factors for weak transitions dominate the uncertainties.
The $(p,p^\prime)$ spectra are background-free up to the neutron threshold and the background due to quasifree scattering is small approaching 2~mb/(sr MeV) \cite{hau91} at higher excitation energies (cf.\ Fig.~\ref{fig:Peak-Markierung}).
Here, the limits come from the sensitivity of the MDA. 
In passing, we note that seven further potential M1 candidates in the $(e,e^\prime)$ data are quoted in Ref.~\cite{ste84}.
However, in the classification scheme introduced in Ref.~\cite{sob85} these fall into lower probability categories.  

\begin{figure}[tbh]
        \centering
 \includegraphics[width=8.6cm]{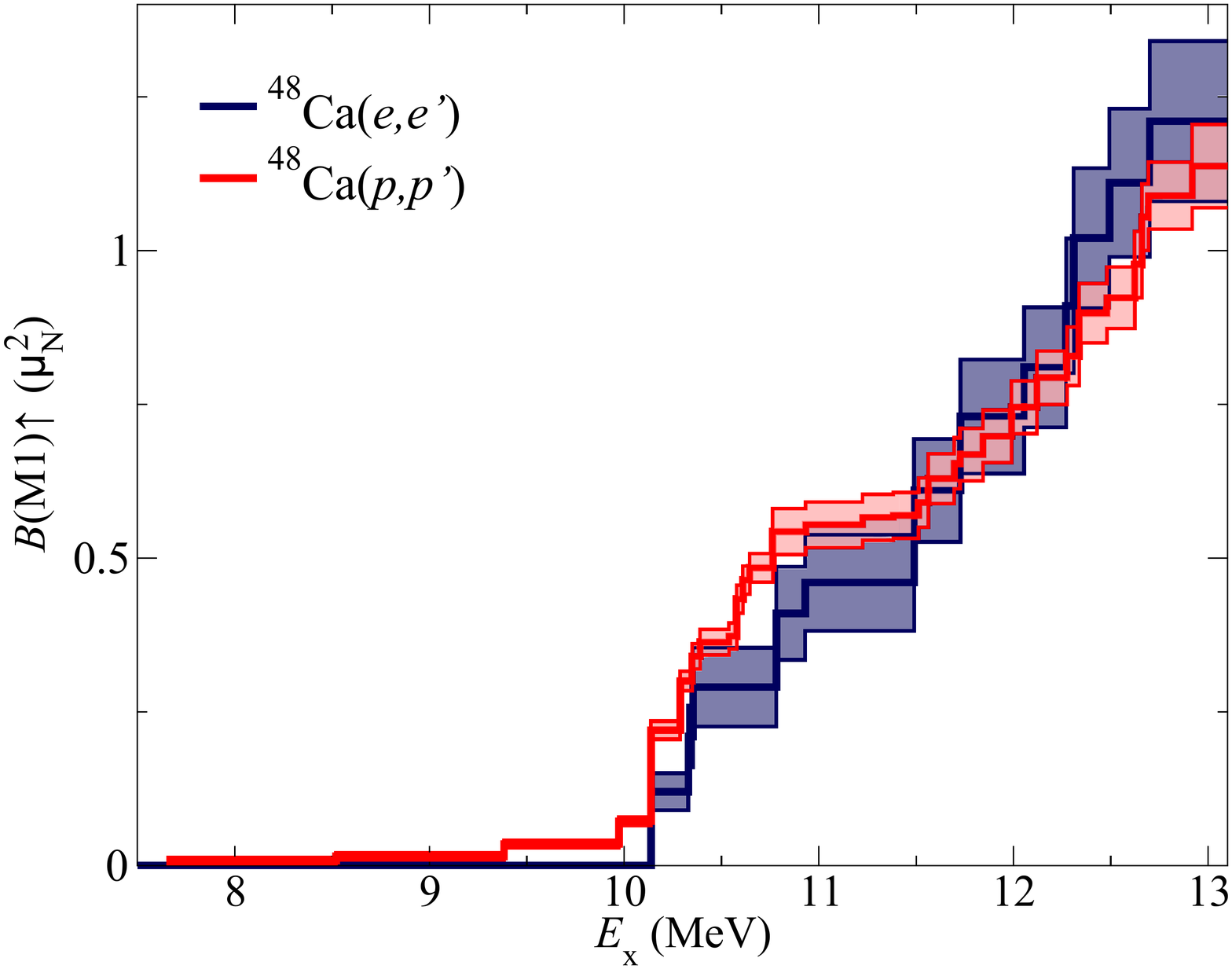}             
\caption{(color online).
Running sums of the  $B(M1)\myuparrow$ strengths in $^{48}$Ca between 7 and 13~MeV (excluding the prominent transition to the state at $E_\mathrm{x}=10.23$~MeV) from the $(p,p')$ reaction (present work, red (light grey) histogram) and the $(e,e')$ reaction (Ref.~\cite{ste83}, blue (dark grey) histogram). 
The bands indicate the experimental uncertainties.}
\label{fig:runningsum}
\end{figure}
Finally, we show a plot of the running sums of the $B(M1)$ strengths from both experiments (Fig.~\ref{fig:runningsum}).
They exhibit a similar slope and agree within error bars except for the region between 10.5 and 11.5~MeV, where the present analysis finds a number of weaker transitions not observed in Ref.~\cite{ste83}.
However, considering that the peaks seen in the spectra of both experiments are near the limits of experimental sensitivity and taking into account the effects which may modify their relative ratio discussed above, the agreement is good.   

\section{Conclusions}
\label{sec:conclusions} 

We have presented a search for $M1$ strength in $^{48}$Ca besides the prominent transition at 10.23~MeV using proton scattering data taken at 295~MeV and very forward angles including $0^\circ$.
The cross sections at $0^\circ$ due to excitation of the spinflip $M1$ mode have been extracted with the aid of a MDA and converted into $B(M1)$ strength with the method outlined in Ref.~\cite{bir16}.
An analysis based on a MDA of individual peaks shows overall good agreement with a study using electron scattering \cite{ste83}.
In detail there are some differences: The $B(M1)$ values from Ref.~\cite{ste83} tend to be higher although they are still consistent within error bars in many cases, and about 50\% more individual transitions are identified in the present data.

The variances between the results from both experiments can be attributed to the different limits of the experimental sensitivity and mechanisms breaking the assumptions made in Ref.~\cite{bir16} for the extraction of electromagnetic transition strengths from the nuclear scattering cross sections, which are aggravated for weak transitions as studied here. 
In particular, contributions from coherent $\Delta L = 2$  and wave function components of the $1^+$ states neglected in the one-phonon approximation of the QPM calculation can modify the $M1$ angular distributions.
Also, the mixing of spin and orbital contributions in the $B(M1)$ strength may play a role.  
It is hard to quantify the related systematic uncertainties because they require explicit models for the wave functions of the ground state and excited states.  
Based on shell-model analyses of these effects in $sd$-shell nuclei \cite{zeg06,fay97} we estimate that they may reach up to 50\% for the weakest transitions studied.

The good correspondence of the total $B(M1)$ strengths deduced from both experiments suggests that there is little additional fragmented strength hidden in the data. 
Accordingly, the quenching factor $g^{\rm IV}_{s,{\rm eff}} \simeq 0.75$ for $M1$ strength deduced in large-scale shell-model calculations \cite{lan04,vnc98} remains, which is comparable to that of GT $\beta$ decay in $fp$-shell nuclei \cite{mar96}.

\begin{acknowledgements}
We are indebted to the RCNP accelerator team for providing excellent beams. 
This work was supported by the DFG under contract No.\ SFB 1245, JSPS KAKENHI Grant No.\ JP14740154, and MEXT KAKENHI Grant No.\ JP25105509. 
\end{acknowledgements}
\end{document}